**Empirical normal intensity distribution for overtone vibrational spectra of triatomic molecules**

Emile S. Medvedev,[1] Vladimir G. Ushakov,[1] Eamon K. Conway,[2,3] Apoorva Upadhyay,[3] Iouli E. Gordon,[2] and Jonathan Tennyson[3]

[1]*The Institute of Problems of Chemical Physics, Russian Academy of Sciences, Prospect Akademika Semenova 1, 142432 Chernogolovka, Russia*
[2]*Atomic and Molecular Physics Division, Harvard & Smithsonian | Center for Astrophysics, 60 Garden St., Cambridge, MA 02138, USA*
[3]*Department of Physics and Astronomy, University College London, London WC1E 6BT, UK*

**Abstract**. Theoretical calculations are contributing a significantly higher proportion of data to contemporary spectroscopic databases, which have traditionally relied on experimental observations and semi-empirical models. It is now a common procedure to extend calculated line lists to include ro-vibrational transitions between all bound states of the ground electronic state up to the dissociation limit. Advanced *ab initio* methods are utilized to calculate the potential energy and dipole moment surfaces (PESs and DMSs), and semi-empirical PESs are then obtained by combining *ab initio* and experimental data. The objective is to reach high accuracy in the calculated transition intensities for all parts of spectrum, i.e. to increase the predictive power of the model. We show that in order to perform this task, one needs, in addition to the standard improvements of the PES and DMS in the spectroscopically accessible regions, to extend the *ab initio* calculations of the PES towards the united-atom limit along the stretching coordinates. The argument is based on the correlation between the intensities of high-overtone transitions and the repulsive potential wall that has previously been theoretically established for diatomic molecules and is empirically extended here to linear and nonlinear triatomic molecules. We generate partial line lists for water and ozone, and together with an already available line list for carbon dioxide, we derive the normal intensity distribution, which is a direct consequence of this correlation. The normal distribution is not an instrument to compute highly accurate intensities, rather it is a means to analyze the intensities computed by the traditional methods.

**1. Introduction**

Knowledge of the spectroscopic parameters of high-overtone transitions of water vapor is important for atmospheric studies [1]. While the intensity of these transitions is quite weak, the increased density of states results in observable contributions at some visible and UV wavelengths in the terrestrial atmosphere. For instance, the spectral region near 450 nm that is dominated by $7v_1$ band is targeted by the upcoming TEMPO mission for monitoring of the water cycle in the terrestrial atmosphere [2]. The same mission will be monitoring the ozone bands and atmospheric pollutants including $NO_x$, $H_2CO$, and BrO, further in the UV. For proper retrieval of these trace gases one has to correctly account for the interference of the water vapor absorption [3]. Experimental measurements of these weak transitions are difficult to do accurately and are therefore scarce and controversial; theoretical calculations offer an attractive alternative.

While the high-overtone transitions of $CO_2$ are not easily observable in the terrestrial atmosphere, they are routinely observed in the atmosphere of Venus (which is dominated by carbon dioxide) and have been observed for almost a century [4]. Therefore, in order to properly model the spectra of the Venusian atmosphere, it is important to accurately know the parameters of such transitions.

To study the weak transitions, the Normal Intensity Distribution (NID) may serve as a useful tool. Originally, the NID was analytically established as a law, the NIDL, for high-overtone vibrational transitions in diatomic molecules and quasi-diatomic local vibrations in polyatomic molecules [5-8]. The validity of the NIDL was verified by experimental and theoretical data in a variety of diatomic molecular systems (see review [9] and references therein). Recently, application of the NIDL as a tool to control the accuracy of the calculated weak intensities in CO has been demonstrated [10]. In this paper, the NID will be empirically established for triatomic molecules.

An important result of the theory is the so-called Effect of the Repulsive Wall (ERW), according to which the rate of decrease in the vibrational band intensity with increasing energy over the full range of transitions up to dissociation mainly depends on the steepness of the repulsive branch of the potential-energy function and only slightly on the dipole-moment function. The ERW implies that intensities of weak transitions are controlled by the behavior of the potential in the repulsive region where the wave function is very close to zero. This appears to be counterintuitive and has been the subject of strong objections in the past [11, 12]. In fact, the origin of the ERW is in the underlying assumption that both the potential energy surface (PES) and dipole moment surface (DMS) are analytical functions, i.e. they and their derivatives of all orders are continuous functions of the inter-atomic separation. According to the well-known mathematical theorem [13-15] any change of the analytical function in the repulsion region will inevitably result in its change over the entire complex plane of the independent variable including near equilibrium, where the wave function has non-negligible amplitude.

Intuitively, the ERW should also apply to polyatomic molecules as their spectra in high-energy regions are well-represented by the quasi-diatomic local vibrations along the stretching coordinates [16-20]. However, there is currently no theoretical NID for polyatomic molecules because the problem is significantly more complex to solve than for diatomic molecular systems. Nevertheless, if the ERW does exist, it should manifest itself as the NID. Therefore, we will derive an empirical NID for triatomics using a large amount of spectroscopic data available for $H_2O$, $O_3$, and $CO_2$, and subsequently investigate how the basic properties of the diatomic NIDL manifest themselves in these triatomic molecules.

The modern trend for the spectroscopic databases such as HITRAN [21] (and especially its high-temperature analogue HITEMP [22]), ExoMol [23], TheoReTs [24], and others is to include calculated transition intensities down to very low values as these become important for high temperature studies. Thus, Polyansky and coworkers calculated the complete line list for water termed 'POKAZATEL' [25], which includes transitions between all bound energy levels of $H_2^{16}O$ up to dissociation [3]. Remote atmosphere sensing for $CO_2$ now requires spectroscopic parameters to be known to very high accuracy, up to 0.5% and better [26]. It is noted in Ref. [27] that "…calculations are increasingly competitive with measurements or, indeed, replacing them and thus becoming the primary source of data on key processes". The latest edition of the HITRAN spectroscopic database [21] contains a significant number of lines of water vapor and carbon dioxide where measured transition intensities featured in earlier editions have now been replaced with *ab initio* calculations. One system for which this is particularly true is $CO_2$; a recent study on $CO_2$ from NASA Ames [28] commented: "…further improvements require sub-percent or sub-half-percent accurate experimental intensities… it is clear that not only must extra care be taken in constructing the DMS in order to reach 1% accuracy, but the Ames-2 PES might also require additional work, especially when theoreticians have started working diligently and ambitiously towards sub-percent or even sub-half-percent accuracy for intensity predictions… **the better Ames-1 positions do not lead to better Ames-1 intensities**." The latter, highlighted statement has direct bearing on the ERW since the weak intensities are sensitive to small changes in the repulsive potential even when the respective energy levels remain nearly unaffected. Therefore, **if such high precision is required for calculated transition intensities, one needs to extend the repulsive wall of the *ab initio* potential up to very high energies**, as we have done for CO [29]. Note that using *ab initio* data to avoid unphysical behavior of the potential at short inter-atomic separations has already become common practice [30, 31].

In Sec. 2, we briefly review the theoretical results for diatomic molecules, and then, in Secs. 3-5, we show how the NID works in water, ozone, and carbon dioxide. The conclusions are summarized in Sec. 6. Manifestation of the NID implies that the intensities of the one-photon transitions in the middle- and high-energy parts of spectrum depend on the repulsive potential. Therefore, in order to increase the accuracy of the calculated intensities of yet unobserved ro-vibrational transitions using model PES and DMS, one has to extend *ab initio* calculation of the model PES into the repulsive region far from equilibrium where no spectroscopic information is available. In other words, the standard procedures of improving the PESs and DMSs should be refined to include these data in the theoretical models. The NID is not intended as a tool to calculate intensities with higher accuracy than before. It is helpful for the analysis of the data obtained by solving the Schrödinger equation with the model PES and subsequent calculations of the transition probabilities with the model DMS.

## 2. Review of the NIDL and ERW for diatomic molecules

The NIDL states that logarithm of the intensity, or the oscillator strength, $f_{nm}$, of a transition between vibrational states $n$ and $m$ is a linear function of the square root of the upper-state energy,

$$\log f_{nm} = \text{constant}_m - a\sqrt{E_n/\omega}, \qquad (1)$$

which is valid when the energy of the upper state $n$ greatly exceeds the characteristic vibrational frequency of the molecule,

$$E_n \gg \omega. \qquad (2)$$

The constant term in Eq. (1) depends on state $m$. If condition (2) is also satisfied for the lower level $m$, the NIDL then takes the form

$$\log f_{nm} = \text{constant} + a\left(\sqrt{E_m/\omega} - \sqrt{E_n/\omega}\right), \qquad (3)$$

where the constant is independent of both $n$ and $m$ and **$a$ is the same as in Eq. (1)**.

The relative oscillator strengths for transitions from a common upper level $n$ to two lower levels $m_1 > m_2$ are obtained from Eqs. (1) and (3), and we can obtain the formula

$$\log\left(f_{nm_1}/f_{nm_2}\right) = \text{constant}_{m_1 m_2}, \qquad (4)$$

if condition (2) is not fulfilled for both lower levels. When condition (2) is fulfilled for the lower levels too, Eq. (3) leads to

$$\log\left(f_{nm_1}/f_{nm_2}\right) = a\left(\sqrt{E_{m_1}/\omega} - \sqrt{E_{m_2}/\omega}\right). \qquad (5)$$

These equations are useful for treating emission spectra where the populations of the upper states are often unknown. Equations (4) and (5) state that **the intensity ratio is independent of the upper-state energy $E_n$**. When both lower states are high enough to fulfil a condition similar to Eq. (2), the NIDL abscissa in Eqs. (3) and (5) is the square-root difference and **the NIDL in Eq. (5) is a straight line going through the coordinate origin**.

It is important to remember that all of the above NIDL equations are approximate, and therefore the deviations from the NIDL predictions can be significant in absolute value. For instance, the intensity ratio predicted by Eq. (5) to be independent of $E_n$ can actually vary by one-two orders of magnitude over the extended energy ranges achievable in modern calculations (see below). Nevertheless, these variations are still small compared to the variations of the absolute oscillator strengths over the same energies, which is in line with the predictions of the NIDL.

For a diatomic, parameter $a$ is inversely proportional to the steepness of the repulsive wall of the potential, hence the ERW occurs with the consequence that the rate of major intensity decrease with increasing energy depends only on the repulsive branch of the potential [9]. The dipole moment function, as well as the attractive branch of the potential, affect the "constant terms" that are in fact slowly varying functions of the quantum numbers $n$ and $m$. The ERW and the NIDL of Eq. (1) are illustrated in Figs. 1 and 2.

The left panel in Fig. 1 compares a distorted Morse potential (DMP) with a reference Morse potential (RMP) as functions of displacement from the equilibrium position, $q = r - r_e$; $\alpha$ is the Morse range parameter. The parameters are selected such that the attractive branches are matched but the repulsive ones are not within the strongly repulsive region up to and far above the dissociation limit. The right panel shows logarithm of the oscillator strength as a function of the square root of the upper-state energy in units of vibrational quantum (points) and the respective NIDLs (lines) calculated for $n \leftrightarrow 0$ transitions with the DMP and RMP potentials and a common dipole function. It is seen that the NIDL is obeyed for overtone transitions ($n > 1$), whereas for the fundamental transition ($n = 1$) the NIDL does not apply. It is also seen that the steeper RMP has the smaller NIDL slope, which confirms the above-mentioned correlation between the NIDL and the repulsive branch.

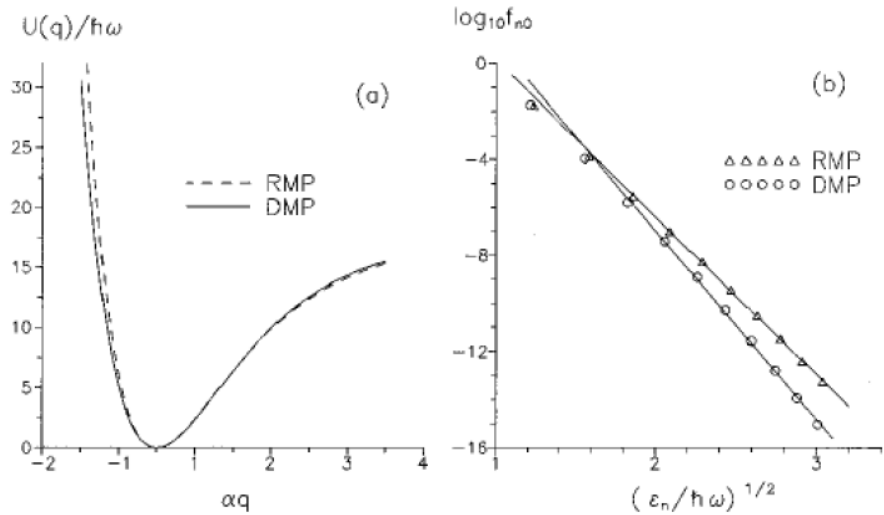

**Fig. 1**. Model potentials to illustrate the ERW (see text). Reproduced from Ref. [32], with the permission of AIP Publishing.

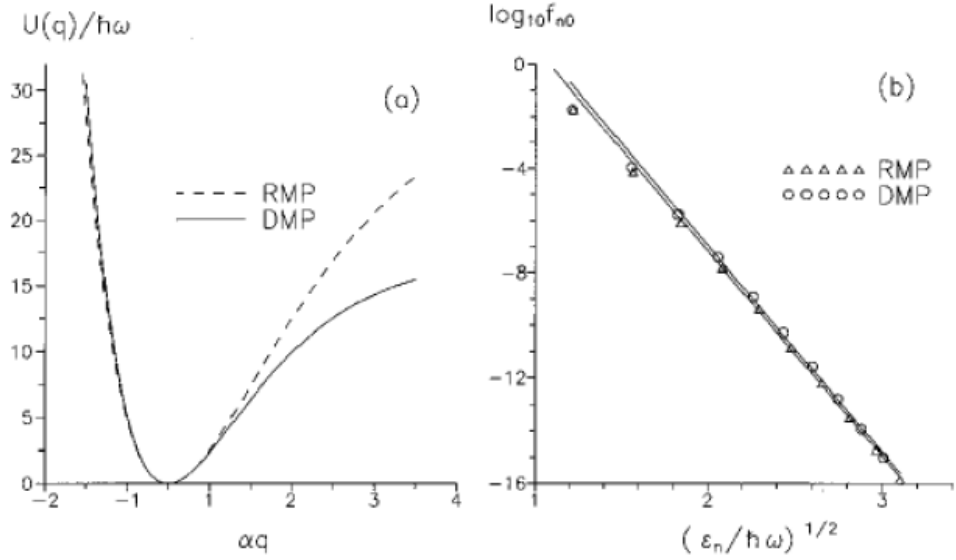

**Fig. 2**. The same potentials as in Fig. 1 but with different parameters (see text). Reproduced from Ref. [32], with the permission of AIP Publishing.

In Fig. 2, the parameters are such that the repulsive branches are matched up to high energies above the dissociation limit but the attractive ones are strongly divergent. The NIDL is again obeyed for both potentials but the slopes are identical. If one considers RMP and DMP as the "true" potential and the "model", respectively, one can say that the model correctly reproduces the pace of the intensity decrease, which is important for achieving high accuracy of the calculations. Yet, the differences between two sets of points remain significant, they can be decreased by proper adjustment of the attractive potential and the dipole moment. This task is simplified here because the differences do not increase for higher transitions as is the case of Fig. 1.

The final result is that the **repulsive branch affects the intensities much more strongly than the attractive branch**. Note that this is valid only for potential and dipole functions that are both analytic functions of the displacement, i.e. they and their derivatives of all orders are continuous. If, for example, the potential abruptly changes its steepness due to a jump in the first derivative at a point in the repulsive region where the wave function has near-zero amplitude, nothing changes in the intensities. Conversely, any analytic perturbation to the analytic potential function always affects the entire range of $q$ including the ranges where the wave-function amplitude is non-zero.

The NIDL equations, Eqs. (1), (3), and (5), were validated for the experimental emission spectra of HF, DF [33], and OH [34] in Refs. [5] and [35]. In particular, the OH data shown in Fig. 7 of Ref. [5] and Fig. 3 of Ref. [35] clearly confirm two predictions of Eq. (5). First, the vertically arranged sets of points at some abscissae correspond to fixed $m_1$ and $m_2$ but varying $n$, and the relative oscillator strength only weakly depends on $n$. Second, the data fall onto a straight line with slope $a = 5.54$ going through the coordinate origin. The line lists for HF and DF as well as HCl and its isotopologues included in the HITRAN2012 database [36] employed semi-empirical calculations by Li *et al.* [37, 38]. These calculations originally extended to very high values of Δv, the change in the vibrational quantum number, and in the process of validating data for the HITRAN2016 edition it was discovered that the intensities of the transitions with Δv > 10 for HF and Δv > 8 for HCl were not following the NIDL line and were considerably overestimated due to numerical issues similar to those discovered in the work on CO [10]. These hydrogen halide line lists were therefore truncated in HITRAN2016.

The examples shown in Figs. 1 and 2 are very crude illustrations of the ERW because the potentials and their respective energy levels differ appreciably. A more realistic example is provided by our recent work on CO, **Meshkov19** [29], for which a new semi-empirical potential was constructed. Compared to the empirical potential of **Coxon04** [39], our potential has the correct short-range behavior whereas the empirical one has a maximum at short inter-atomic separations. The difference between the potentials within the spectroscopically accessible region v < 10 is extremely small so that the energy levels predicted by both models only differ within their experimental uncertainties. The ERW under consideration is characterized by the following CO data, see Fig. 3.

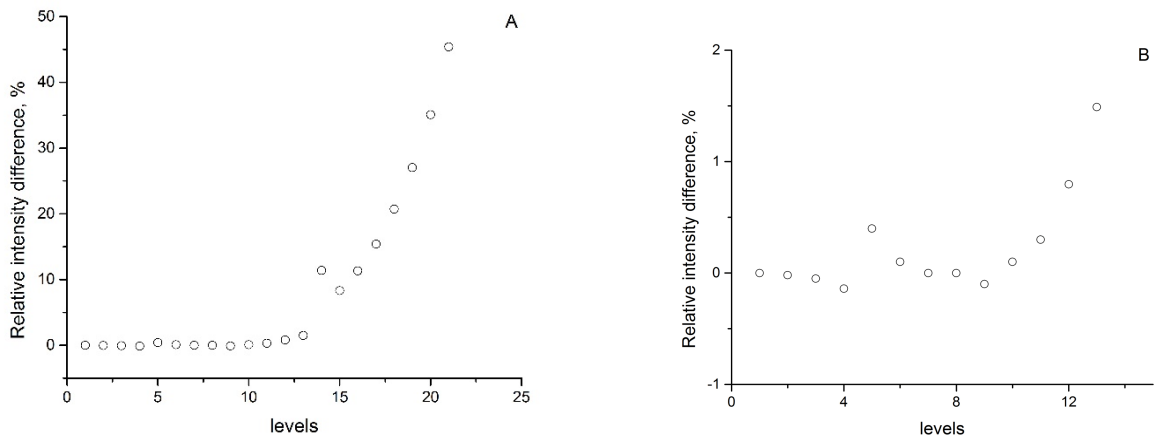

**Fig. 3**. The relative intensity difference between two CO potentials, **Meshkov19** minus **Coxon04**, which have nearly identical energy levels but strongly divergent repulsive branches. Note the different ordinate scales between two panels. Panel B shows that the anomaly at v = 5 (see Sec. 5) is much more sensitive to the potential than other transitions to the spectroscopically accessible levels v < 10.

The intensities calculated with these two potentials and a common dipole moment function (the empirical DMF from [10]) diverge at $n > 25$ (not shown). As seen in Fig. 3, for $n = 20$ the intensities differ by 40% and for $n = 15$ they differ by 10%. The most dramatic changes occur outside the observable range of transitions. Within the $n = 5$-11 transitions, the intensities differ by no more than 0.5%. Is this significant? This depends on our requirements. If, as stated in the Introduction, reaching such high accuracy is necessary, we must develop potentials that extend high enough into the repulsive region. The current PES and DMS models are based on thousands of observed lines whereas the resulting line lists contain millions of lines in all parts of spectrum. Figure 3 shows that accuracy of the calculated intensities, both at high and low energies, depends not only on the experimental and theoretical data in the spectroscopic region and slightly outside it, e.g. *ab initio* calculations near the dissociation limit, but also on data far above the dissociation limit.

### 3. Water

In order to derive the empirical NID, we will plot the calculated Einstein *A* coefficients, $A_{nm}$, the oscillator strengths, $f_{nm}$, or the transition dipole-moment squares, $R_{nm} = |d_{nm}/\text{debye}|^2$, in the NID coordinates, $\log(A, f, \text{or } R)$ vs $\sqrt{E_n/\omega}$. They are connected by the formulae

$$A_{nm} = (0.66703 \times \text{Hz})(\tilde{\nu}_{nm}/\text{cm}^{-1})^2 f_{nm} = (3.136189 \times 10^{-7}\,\text{Hz})(\tilde{\nu}_{nm}/\text{cm}^{-1})^3 R_{nm}, \quad (6)$$

where $\tilde{\nu}_{nm}$ is the transition wavenumber in $\text{cm}^{-1}$. Note that the NID stems from the properties of *R* whereas the additional frequency factors entering the other quantities essentially do not affect the NID, they just slightly alter the NID slope, as we will see below.

The NID equations, Eqs. (1), (3)-(5), are now rewritten in a form suitable for polyatomic molecules whose states have no vibrational labelling at high energies,

$$\log f_{E'E''} = \text{constant}_{E''} - a\sqrt{E'/\omega}, \qquad E'' < 2\omega, \quad (7)$$

$$\log\left(f_{E'E_1''}/f_{E'E_2''}\right) = \text{constant}_{E_1''E_2''}, \qquad E_2'' < E_1'' < 2\omega, \quad (8)$$

$$\log f_{E'E''} = \text{constant} + a\left(\sqrt{E''/\omega} - \sqrt{E'/\omega}\right), \qquad E'' \geq 2\omega, \quad (9)$$

$$\log\left(f_{E'E_1''}/f_{E'E_2''}\right) = a\left(\sqrt{E_1''/\omega} - \sqrt{E_2''/\omega}\right), \quad E_1'' > E_2'' \geq 2\omega, \quad (10)$$

where $E'E''$ replaces $E_n, E_m$ and condition (2) is reduced to $E \geq 2\omega$. The prime and double prime refer to the upper and lower states, respectively.

At this junction, we turn to the line lists containing data on the ro-vibrational transitions. As follows from the above considerations, the NID relates to purely vibrational transitions, therefore we will consider only transitions with minimum rotational contributions (*J* = 0 and 1). Only one-photon absorption transitions will be discussed.

Recently, Polyansky *et al.* [25] created the semi-empirical POKAZATEL PES for water and calculated the line list using the DMS from Lodi *et al.* [40]. Shortly after this, an improved DMS termed CKAPTEN was developed by Conway *et al.* [41]. Here, we generate two partial line lists with the POKAZATEL PES and the above two DMSs. The calculations are performed with the cutoff $A_{E'E''} > 10^{-15}$ s$^{-1}$ and only for $J = 0$ or 1 since the NID relates to purely vibrational transitions, i.e. transitions near the band origins. These line lists are provided in the supplementary material.

Figure 4 presents our data recalculated as oscillator strengths using Eq. (6). The characteristic frequency of the OH radical, $\omega = 3735$ cm$^{-1}$ [42, 43] (different by 3 cm$^{-1}$ in Ref. [44]), was used to calculate the NID abscissa. The NID lines are calculated as least-squares fits to the points shown.

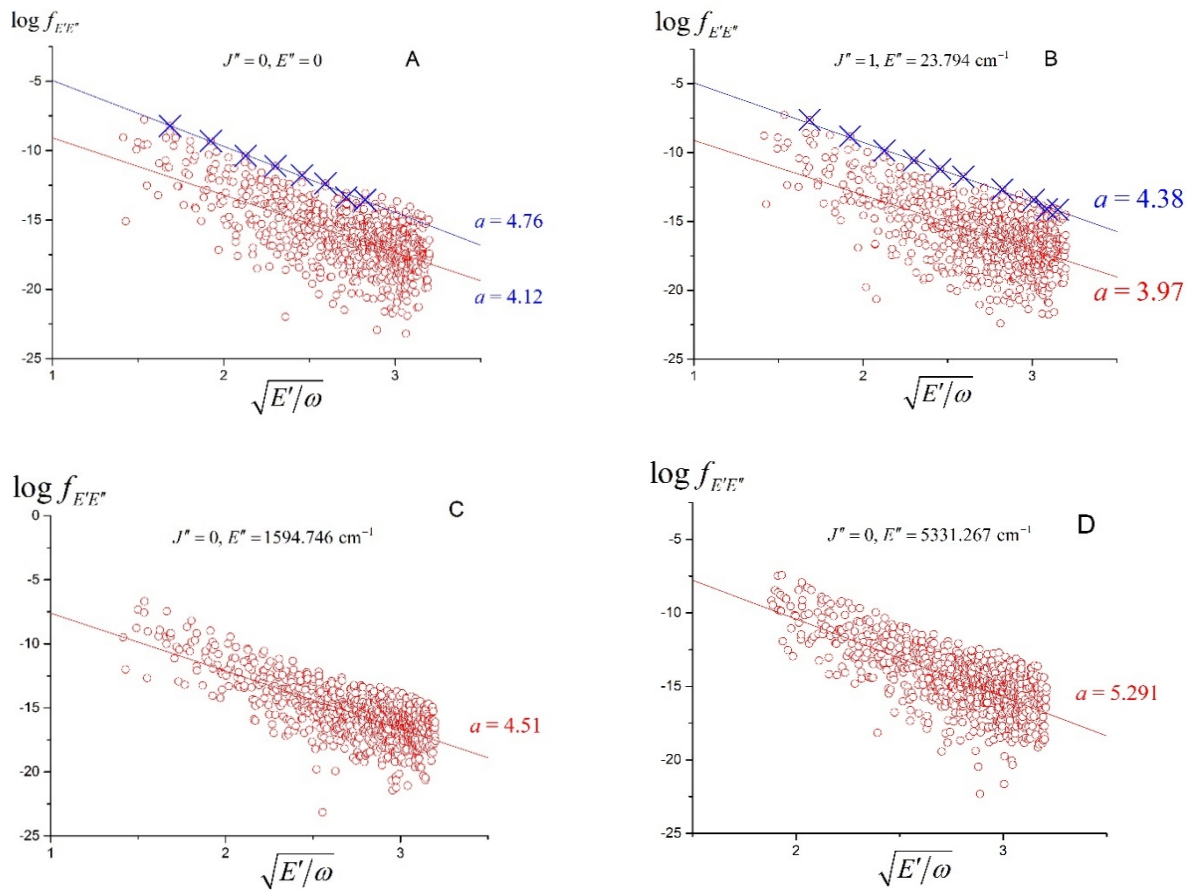

**Fig. 4**. Logarithm of the oscillator strengths of the allowed H$_2^{16}$O transitions from all upper states with $E' > 2\omega$ to a fixed lower state. Panel A, transitions between $J' = 1$ and $J'' = 0, E'' = 0$; panel B, transitions between $J' = 0$ and $J'' = 1, E'' = 23.794$ cm$^{-1}$; panels C and D, "hot" transitions ($T \approx 2300$ and 7600 K, respectively) starting at excited lower states. Circles, data of the present study calculated with POKAZATEL PES and CKAPTEN DMS; crosses, selected quasi-diatomic transitions to the local-mode-type upper states; lines, the respective NID plots, Eq. (7), with indicated slopes $a$. (Colour online).

It is evident that the NID is obeyed up to the dissociation limit for all four lower states selected. The NID slope depends only weakly on $E''$, as predicted by Eq. (7), and also on $J$ at low $J$ since the NID is a property of the vibrational transitions. However, there is a large scatter of points around the NID, therefore additional verification is required. One would not expect the NID to be followed with the same precision as for diatomic molecules because of the bending mode that does not obey Eq. (1). Another reason for the large scatter is that the transitions with the excitation transfer between two OH stretches (the so-called intensity "stealing" from strong transitions by weak ones) also do not obey the NID. Nevertheless, it should be noted how closely the NID linear dependence is followed at the top edge of the scatter. Another interesting feature is that the NID slope is relatively close to the value of 5.54-5.88 calculated for the OH radical from experimental intensities [35].

The features mentioned above deserve discussion in more detail. First of all, since the NID is a property of diatomic molecules, it is expected to manifest for transitions to those states that are characterized by large-amplitude motions of one atom while the two remaining atoms perform low-amplitude vibrations around equilibrium. These states are better described as the local modes rather than the normal ones, which is confirmed by the analysis of the wave functions [19]. Plots of wave function amplitudes of stretching states with no bend excitation show the strongest extrema at configurations where either $R_1 \gg R_e, R_2 \approx R_e$ or $R_1 \approx R_e, R_2 \gg R_e$ for states with both symmetric and antisymmetric normal-mode labels. These configurations clearly correspond to the 1D motion of one H atom with respect to the remaining OH group. Therefore, the states under considerations are quasi-diatomic and transitions to them from the lower states should obey the NID.

Secondly, any breakdown of the NID leads to an intensity decrease because it is much less probable for the photon energy to be shared between two or three degrees of freedom than to be absorbed by a single one. Therefore, the NID relates to the transitions with the highest intensities. Following this logic, we made selections of transitions using local-mode notation $(n, m)^\pm v_2$ where $n$ and $m$ are expressed in terms of the normal-mode quantum numbers $v_1$ and $v_3$ of the symmetric and antisymmetric stretches, respectively [45, 46]. Specifically, states $(n, 0)^\pm 0$ were considered. The "±" states with a given $n$ are nearly degenerate, the small splitting being due to the tunneling transitions [20], in which the excitation of $n$ quanta in one OH bond is transferred to the other bond. One of these states has a large matrix element for transition to a lower state obeying the NID whereas the transition from the other is much less probable. The results of this analysis for the transitions shown in Figs. 4A and 4B are presented in Tables S1 and S2 of the supplementary material. In both tables, transitions to the excited "−" states were selected, and their respective points are shown by crosses in Figs. 4A and 4B. Inspection of the tables shows that the "±" states are indeed nearly degenerate and the "−" transitions are the strongest ones, that the latter closely follow the NID, and that the NID slope for this particular subset of data is close to that which characterizes the whole data set. The physical meaning of the slope is that it is inversely proportional to the steepness of the repulsive part of the diatomic potential in the deeply repulsive region far from equilibrium [9]. Therefore, for a quasi-diatomic HO-H bond vibration, the repulsive branch should be similar to that in O-H, hence the slopes are also expected to be similar. This is indeed true for the data in Fig. 4.

Panels C and D in Fig. 4 show that the NID is also obeyed for the transitions involving excited lower states. Again, the NID is clearly seen at the top edge of the scatter and its slope is approximately the same as for the transitions from the ground state.

The next prediction of the theory for diatomic molecules is that the dipole moment only slightly affects the NID slope as compared with the repulsive potential ("mechanical anharmonicity is more important than electrical anharmonicity" [47]). This result is validated in Fig. 5 where the data calculated with a common PES and two different DMSs are shown. The respective NID slopes differ only by 0.16. In Ref. [9], the maximum variation of the slope due to varying dipole moment was estimated to be 0.7. One can argue that the difference between the DMSs used is small, hence the difference between the corresponding intensities should also be small. However, this is only half true: the difference in the intensities is small only at low excitations where experimental information is available and the DMSs are well defined (i.e., near equilibrium). Conversely, at high energies, where experimental and theoretical information on the DMS is not readily available, differences in the calculated intensities can be very large, as is seen in Fig. 5.

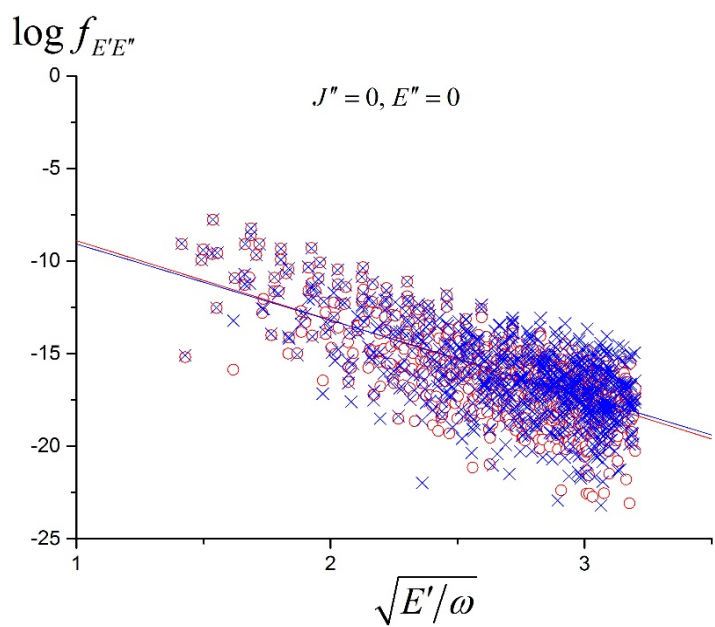

**Fig. 5**. Comparing the oscillator strengths for $H_2^{16}O$ transitions from two line lists calculated with the common POKAZATEL PES and two different DMS. Circles, Lodi *et al*. DMS [40]; crosses, CKAPTEN DMS; lines, the corresponding NID plots, Eq. (7), with slopes $a = 4.28$ and 4.12, respectively. (Colour online).

Further, we show how the prediction of Eq. (8) is fulfilled. Figure 6 plots the squares of the relative transition matrix elements between a common upper state $E'$ and two different lower states $E''_{1,2}$ (left ordinates in two panels for two pairs of the lower levels). In order to minimize the deteriorating effect of the above-mentioned transitions that do not obey the NID, we select only the transitions between the states with assigned local-mode vibrational quantum numbers such that $v'_2 = v''_2 = 0$. For comparison, the intensities of the individual transitions to states $E''_1$ and $E''_2$ are shown (right ordinates). It can be seen that the intensity ratio in two panels varies within a factor of 10 over the wide intervals of $E'$ whereas the absolute transition intensities to states $E''_1$ and $E''_2$ vary by more than five orders of magnitude over the same intervals.

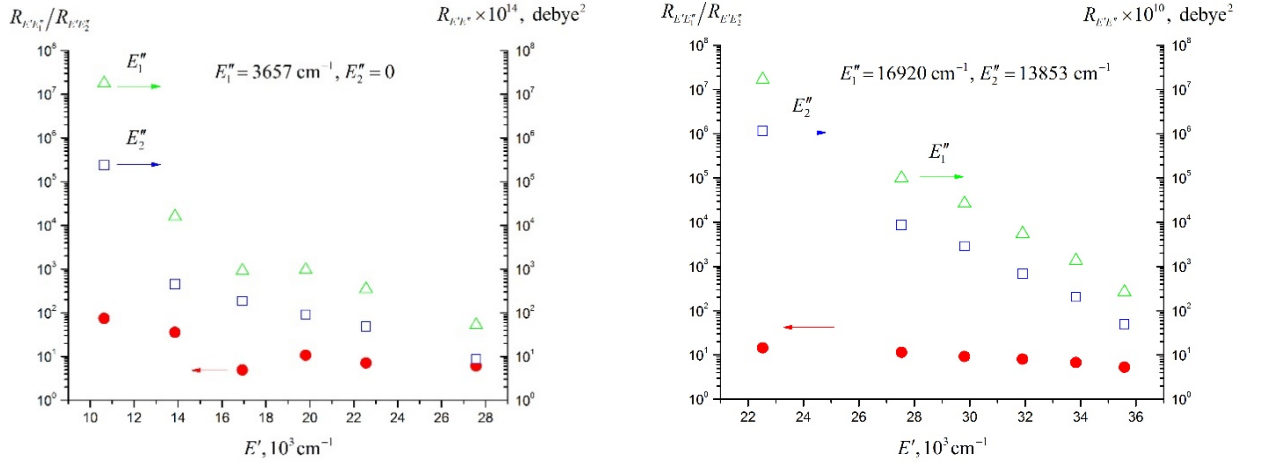

**Fig. 6**. Demonstration of the rule expressed by Eq. (8). The relative intensities (filled circles, left ordinates) are weakly dependent of the upper-state energy $E'$ for two pairs of lower levels, $E_1''$ and $E_2''$, shown at the top of two panels. Empty triangles and squares, absolute intensities of the transitions to states $E_1''$ and $E_2''$, respectively (right ordinates). Arrows point to the left or right ordinate axis relevant to the respective data sets. (Colour online).

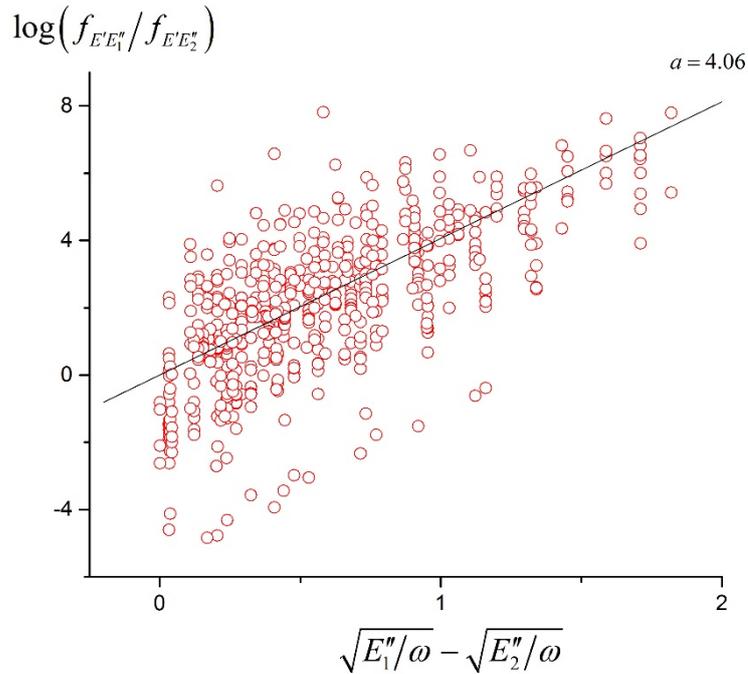

**Fig. 7**. Validation of Eq. (10). Points, data of the present study. Transitions between the states with the assigned local-mode quantum numbers, $v_2$ being unchanged, and $J = 0$ or 1 are shown. Vertical sets of points correspond to transitions from a common upper level $E'$ to two lower levels $E_1'' > E_2''$. Line, the corresponding NID plot with slope $a = 4.06$ found by linear least-squares fitting with fixed zero intercept. (Colour online).

The final test in Fig. 7 relates to Eq. (10). In order to reduce the effect of those transitions that do not obey the NID, we select transitions between the assigned states and include only those for which the number of quanta in the bending mode is unchanged. We see that the NID is again fulfilled, but only qualitatively as the scatter remains very large. In order to better understand the large scatter in the data set, we investigated the NID behavior of the water transitions in more detail. Namely, we considered a selected number of transitions from the ground (0,0,0) state to the normal-mode states ($v_1$,0,0), (0,0,$v_3$), to the local-mode ones ($n$,0)$^+$0, ($n$,0)$^-$0, and to those other states where only one of the vibrational labels changes. Independent of the labelling system used, each such sequence describes transitions with excitation of a specific quasi-1D vibration along some 1D section of the full PES. Such behavior was confirmed by the analysis of the wave functions [19], which clearly demonstrated the 1D character of the highly excited states. If the resulting potential-energy curve has a pronounced repulsive branch, the transition intensities must obey the NID over the entire energy range up to dissociation, and this prediction of the theory is observed (not shown). Thus, the entire sets of points in Figs. 4, 5, and 7 are all composed of sub-sets of points obeying their own NIDs with their own slopes and displacements along the ordinate axis. This is the reason why a large amount of scatter is observed.

We conclude that, since the NID is associated with the repulsive branch of the potential, knowledge of the PES in the repulsive region is important for high-precision calculations of the transition intensities.

## 4. Ozone

Ozone is a very special molecule. Its dissociation limit is so low that the NID has little room to manifest itself, i.e. the number of the high-energy transitions is much smaller than in other triatomics that dissociate at much higher energies, such as water. Despite this, it is important to investigate this type of molecule.

There have been many measurements and calculations of ozone vibration-rotation transitions (see for instance [48-51]). For our purposes, we calculate the $J'= 0,1$ to $J''= 1,0$ transition intensities, as well as the vibrational band origins, for all vibrational states up to dissociation using the PES and DMS from Polyansky $et\ al.$ [52]. As in the case of water, the calculations are performed with the cutoff $A_{E'E''} > 10^{-15}$ s$^{-1}$. We make assignments by comparing our band origins with those of Zúñiga $et\ al.$[53]. The results are shown in Fig. 8, where the abscissa values are calculated with ω = 1580 cm$^{-1}$, the fundamental frequency of the ground-state oxygen $O_2$ [43, 44]. The numerical data are given in the supplementary information.

While the points in Fig. 8 show large scatter, the upper edge of the scatter follows the NID with much higher precision. As in the case of water, we tried to select the transitions to the upper 1D-like states. However, in view of the well-known difficulties with the labelling of the highly excited states [53], the selection of states was not as straightforward as for water.

First, we select transitions to the states labelled as ($v_1$, 0, $v_3$) with $v_3$ odd to obtain the sequence of the "–" states that should mimic a quasi-1D stretch motion. Ten such transitions are identified in the low-energy region, nine of them (i.e. excluding one outlier) are shown by crosses, and they indeed follow the NID along the upper edge of the scatter in Fig. 8.

Despite the fact that the upper NID extends up to the dissociation limit, no "–" states with normal/local-mode labels are found in the uppermost energy region where the assignments become progressively more difficult. Therefore, we tried the second method based on the physical considerations, a kind of "physical" labelling. Like water [17], doublets at very high energies of the nearly degenerate levels must exist in ozone as well. Therefore, we searched pairs of closely spaced levels, located within 10 cm$^{-1}$, with a large difference in the transition intensities.

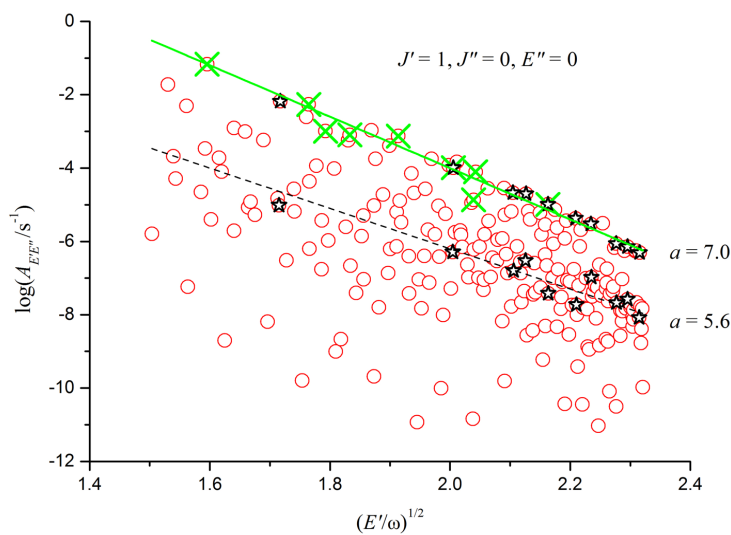

**Fig. 8**. Logarithm of the Einstein $A$ coefficients of the $^{16}O_3$ allowed transitions from all upper states $J' = 1$ to the $J'' = 0$, $E'' = 0$ ground state. Circles, data of the present study; crosses and stars, selected quasi-diatomic transitions (see text); lines, the respective NID plots, Eq. (7). (Colour online).

Some of such pairs found are shown by stars in Fig. 8, with the intensity difference within the pairs being two orders of magnitude and more. The vibrational parts of the wave functions of the doublet components as well as two components of the dipole moment in the molecular plane are of the "±" symmetry types. It turns out that transitions from the ground state to the excited "+" states are strongly suppressed compared to the excited "–" states. This feature was further verified by placing the "±" labels to all levels even where no normal/local-mode labelling was possible. Figure S1 of the supplementary material shows that, on average, the "–/+" transitions are indeed much stronger than the "+/+" ones. It is also worth mentioning that the strong and weak $\Delta J = 1$ transitions considered in this paper obey the rules $\Delta K_a = 0$ and 1, respectively, in both ozone and water.

It is remarkable that, in Fig. 8, the stronger transitions closely follow, up to dissociation, the upper NID built on the lower-energy transitions. It should be noted that the strong transitions in the high-energy region are not necessarily due to purely stretching vibrations, the bending mode can contribute as well. In the highly excited molecule with three identical atoms, the difference between stretch and bend becomes vague. It is noticeable that the weaker transitions shown in Fig. S1 follow their own NID too because they are also due to 1D-like motions. It should be mentioned that there are a lot of other pairs of transitions (not shown) that have comparable probabilities for their two members and originate from an occasional coincidence of the energies in the dense spectrum and that there is no means to separate out the occasional coincidences except for our "physical" principle. The data for crosses and stars are given in the supplementary material.

Despite the above-mentioned difficulties with ozone as a weakly bound molecule, we believe that the arguments presented are sufficient to state that the ozone transitions obey the empirical NID just as more "typical" triatomics like water.

## 5. Carbon dioxide

The most recently available theoretical data extending to the higher vibrational levels for $CO_2$ come from Huang *et al.* [28]. Their calculations are shown in Fig. 9 along with the corresponding NID. The vibrational frequency is $\omega = 2170$ cm$^{-1}$ [42, 43]. The NID slope is 5.80, which can be compared with the values of 7.2-8.1 estimated for CO in Table 1 of Ref. [54]. Several factors can affect the slope, the most important being the steepness of the repulsive branch for OC-O vibration and the effective mass for this movement. Since the cutoff of the Huang *et al.* data was put on the intensities rather than on the Einstein $A$ coefficient, the NID slope is also slightly affected by the populations. The scatter is again large due to the transitions that do not obey the NID. Nevertheless, just as for water and ozone, the NID behavior of the top edge of the scatter is clearly seen.

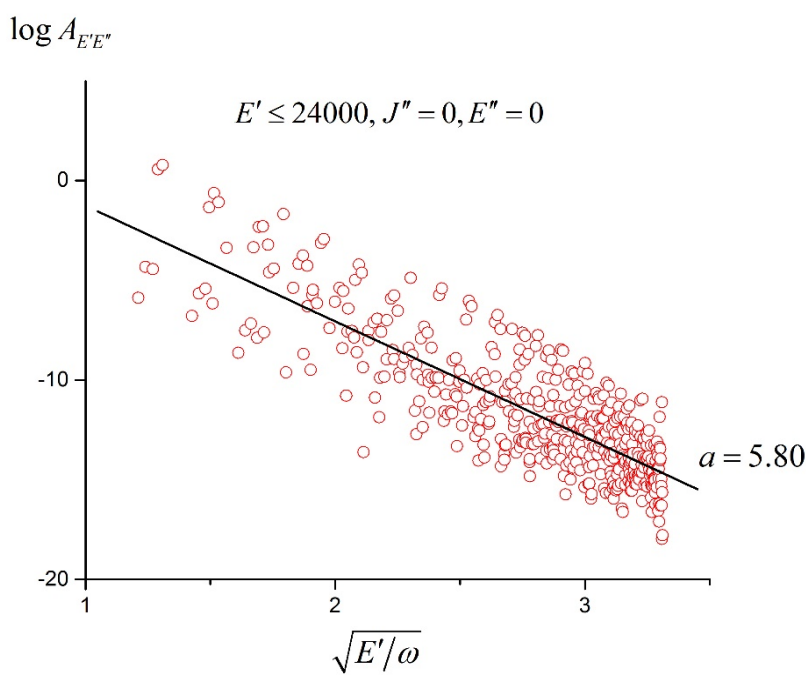

**Fig. 9**. The calculated Einstein $A$ coefficients for $^{12}C^{16}O_2$ (Ref. [28]) in the NID coordinates (circles) and the NID plot (line). (Colour online).

**6. Discussion and conclusion**

The Einstein $A$ coefficients and the oscillator strengths for water and ozone $J$ = 1-0 and 0-1 transitions were recalculated. For water, we used the recently developed POKAZATEL potential-energy surface [25] and two different dipole-moment surfaces [40, 41]; for ozone, we used the surfaces from Polyansky *et al*. [52]. The calculated transition intensities for $CO_2$ were taken from Ref. [28]. Analysis of these data sets confirmed that excitation of stretching motions obey the Normal Intensity Distribution (NID) earlier established for diatomic molecules.

The large scatter of the data around the NID plots is due to those transitions that do not obey the NID. We emphasize that the NID is associated with the stretching potential in the repulsive region. In water, ozone and carbon dioxide, two stretching modes obey the NID, while the bending modes do not. Moreover, transitions that transfer excitation between different modes do not obey the NID either. By selectively choosing transitions to quasi-diatomic states where one atom performs large-amplitude motion and the other two remain near equilibrium, the amount of scatter is greatly diminished.

In the case of diatomic molecules, the use of the word "law" is justified by the facts that the NIDL 1) was derived theoretically; 2) was confirmed by all existing experimental and computational data; and 3) applies to any diatomics without exclusion. Even the anomalies, i.e. isolated bands whose intensities fall well below the NIDL line (as exemplified by CO, see Fig. 8 in Ref. [29]) do not contradict such a classification because there is an underlying physical explanation, in this case interference; therefore, they are not "unphysical" as stated recently [55]. In the case of triatomic molecules, the NID has not been derived theoretically but by analogy with diatomic systems. Still we believe that the data presented in this paper strongly support the NID as a valid empirical relation for triatomic molecules. The NID has proved to be a useful tool for validating the intensities of the high overtone transitions [10].

As follows from the theory [9], the manifestation of the NID implies that the intensities of the weak high-overtone vibrational transitions (and presumably the anomalies [56, 57]) essentially depend on the behavior of the potential-energy function in the deep repulsive region far from equilibrium, in contrast to the intensities of strong (fundamental and low-overtone) transitions. However, if the aim is to reach high accuracy in the calculations, the effect of the repulsive wall must be taken into account even for the spectroscopically important transitions. Therefore, when improving the accuracy of intensity calculations, *ab initio* calculations must extend to regions far from equilibrium, ideally to both the united-atom and separated-atoms limits. This is a necessary requirement, yet of course it is far from being sufficient. It goes without saying that the traditional approach based on improving PESs and DMSs in the spectroscopically accessible regions remain important means toward high-accuracy calculations. However, modelling the potential- and dipole-moment functions with use of such information along with the experimental data is not a straightforward task, as has been recently shown by our experience with CO [29], [58]. In particular, the traditional potential forms based on the generalized Morse function fail, and new forms are required to embrace all the data.

As should be obvious from the above, the normal distribution is not a tool for calculating intensities with high precision, but it is of value as a tool for analysis of the data computed using traditional procedures. The significance of the present findings for the actual calculations stems from the fact that the calculated data greatly overwhelm the experimental data even within the spectroscopically observed regions and therefore make significant contributions to the predictions of various theories based on the published line lists. The inclusion of the *ab initio* data for the repulsive potential into the model of the potential surface is important for enhancing the predictive power of the theoretical models with respect to the transition intensities in the middle- and high-energy regions of the spectrum.

**Conflicts of interest**

There are no conflicts to declare.

**Acknowledgements**

The work was performed in accordance with the state task (to ESM and UVG), state registration No. 0089-2019-0002. JT and AU thank the UK Natural Environment Research Council (NERC) for funding through NE/T000767/1 and other grants.